\documentclass[twocolumn]{aastex631}

\begin{document}


\title{Reliable Detections of Atmospheres on Rocky Exoplanets with Photometric JWST Phase Curves}

\author[0000-0002-6893-522X]{Mark Hammond}
\affiliation{Atmospheric, Oceanic, and Planetary Physics, Department of Physics, University of Oxford, Parks Rd, Oxford OX1 3PU, UK}
\author[0000-0003-1521-5461]{Claire Marie Guimond}
\affiliation{Atmospheric, Oceanic, and Planetary Physics, Department of Physics, University of Oxford, Parks Rd, Oxford OX1 3PU, UK}
\author[0000-0002-3286-7683]{Tim Lichtenberg}
\affiliation{Kapteyn Astronomical Institute, University of Groningen, P.O. Box 800, 9700 AV Groningen, The Netherlands}
\author[0000-0002-8368-4641]{Harrison Nicholls}
\affiliation{Atmospheric, Oceanic, and Planetary Physics, Department of Physics, University of Oxford, Parks Rd, Oxford OX1 3PU, UK}
\author[0000-0003-0652-2902]{Chloe Fisher}
\affiliation{Astrophysics, Department of Physics, University of Oxford, Parks Rd, Oxford OX1 3PU, UK}
\author[0000-0002-4671-2957]{Rafael Luque}
\affiliation{Department of Astronomy \& Astrophysics, University of Chicago, Chicago, IL 60637, USA}
\affiliation{NHFP Sagan Fellow}
\author[0000-0003-4143-8482]{Tobias G. Meier}
\affiliation{Atmospheric, Oceanic, and Planetary Physics, Department of Physics, University of Oxford, Parks Rd, Oxford OX1 3PU, UK}
\author[0000-0003-4844-9838]{Jake Taylor}
\affiliation{Astrophysics, Department of Physics, University of Oxford, Parks Rd, Oxford OX1 3PU, UK}
\author[0000-0001-6516-4493]{Quentin Changeat}
\affiliation{Kapteyn Astronomical Institute, University of Groningen, P.O. Box 800, 9700 AV Groningen, The Netherlands}
\affiliation{Department of Physics and Astronomy, University College London, Gower Street, WC1E 6BT London, UK}
\author[0000-0003-4987-6591]{Lisa Dang} 
\affiliation{Trottier Institute for Research on Exoplanets and Département de Physique, Université de Montréal, 1375 Ave Thérèse-Lavoie-Roux, Montréal, QC, H2V 0B3, Canada}
\author[0000-0003-1746-1228]{Hamish C. F. C. Hay} 
\affiliation{Department of Earth Sciences, University of Oxford, S Parks Rd, Oxford OX1 3AN, UK}
\author[0000-0002-1807-4441]{Oliver Herbort}
\affiliation{Department for Astrophysics, University of Vienna, Türkenschanzstrasse 17, A-1180 Vienna, Austria}
\author[0009-0008-2801-5040]{Johanna Teske} 
\affiliation{Earth and Planets Laboratory, Carnegie Institution for Science, 5241 Broad Branch Road, NW, Washington, DC 20015, USA}

\begin{abstract}
The prevalence of atmospheres on rocky planets is one of the major questions in exoplanet astronomy, but there are currently no published unambiguous detections of atmospheres on any rocky exoplanets. The MIRI instrument on JWST can measure thermal emission from tidally locked rocky exoplanets orbiting small, cool stars. This emission is a function of their surface and atmospheric properties, potentially allowing detections of atmospheres. One way to find atmospheres is to search for lower day-side emission than would be expected for a black body planet. Another technique is to measure phase curves of thermal emission to search for night-side emission due to atmospheric heat redistribution. Here, we compare strategies for detecting atmospheres on rocky exoplanets. We simulate secondary eclipse and phase curve observations in the MIRI F1500W and F1280W filters, for a range of surfaces (providing our open access albedo data) and atmospheres on thirty exoplanets selected for their F1500W signal-to-noise ratio. We show that secondary eclipse observations are more degenerate between surfaces and atmospheres than suggested in previous work, and that thick atmospheres can support emission consistent with a black body planet in these filters. These results make it difficult to unambiguously detect or rule out atmospheres using their photometric day-side emission alone. We suggest that an F1500W phase curve could instead be observed for a similar sample of planets. While phase curves are time-consuming and their instrumental systematics can be challenging, we suggest that they allow the only unambiguous detections of atmospheres by night-side thermal emission.
\end{abstract}



\section{Introduction} \label{sec:intro}

The distribution of atmospheres on rocky planets is one of the major unanswered questions of exoplanet science, with wide-ranging implications for the formation pathway, climate variability, and ultimately surface habitability of potentially Earth-like worlds \citep{Wordsworth2022,Lichtenberg2024}. So far, there are no rocky exoplanets with unambiguously detected atmospheres. The James Webb Space Telescope (JWST) can measure thermal emission from rocky planets orbiting smaller, cooler stars, which may contain indications of the presence or absence of atmospheres. The signal-to-noise ratio for these observations is highest in the mid-infrared, around 10 to 15 $\mu$m, which overlaps with a CO$_{2}$ absorption feature. Observations with the F1500W and F1280W photometric filters of the MIRI instrument on JWST \citep{rieke2015mid} (and possibly the MIRI/LRS instrument for hotter planets) therefore have the ability to detect departures from the thermal emission due to a black body planet. These departures could be caused by changes to albedo, surface spectral features, atmospheric spectral features, or atmospheric heat redistribution. These features can have degenerate effects on thermal emission, especially in the bandpasses of individual photometric filters.

Using such observations, STScI is currently implementing a survey of rocky M-dwarf exoplanets with 500 hours of JWST Director's Discretionary Time (DDT), informed by the report from the Working Group on Strategic Exoplanet Initiatives with HST and JWST \citep{redfield2024report}. \citet{redfield2024report}  proposed a programme to observe 15-20 rocky planets with 2 to 15 secondary eclipses each using the MIRI F1500W filter. Its aim is to detect emission lower than expected for a bare-rock planet without an atmosphere, which is proposed to be a signature of atmospheric heat redistribution \citep{koll2019identifying,koll2022scaling}. Detecting atmospheres in this way is proposed as a test of the ``cosmic shoreline'' hypothesis relating to the prevalence of atmospheres on rocky exoplanets \citep{zahnle2017cosmic}. The purpose of this paper is to simulate potential surfaces and atmospheres on a range of the most observable rocky exoplanets, to identify degeneracies and to identify unambiguous signatures of atmospheres. We aim to use these results to inform the planning and analysis of observations of rocky planets with JWST.

The use of MIRI photometric filters to detect signatures of atmospheres on rocky exoplanets was first demonstrated in detail by \citet{deming2009discovery}, which identified the significantly lower emission at $15\mu$m that could be produced by CO$_{2}$ absorption, as well as a correspondingly higher emission at $11.3\mu$m. We follow in particular the work of \citet{hu2012theoretical}, which identified the variety of surface albedos and identifiable spectral features possible for a range of exoplanet surfaces. 

\citet{mansfield2019identifying} simulated the thermal emission observed by the JWST MIRI/LRS instrument for the surface types considered in \citet{hu2012theoretical}, showing how the observed emission depends on the surface type. \citet{mansfield2019identifying} also showed how bare-rock planets can emit more strongly from their day-sides at longer wavelengths than would be expected from their equilibrium temperatures, due to the relatively higher emissivity of the modelled surfaces at longer wavelengths. They suggested that their results ruled out day-side bare-rock emission below a certain value, so could be used as a method of detecting atmospheres on rocky planets. \citet{koll2019identifying} compared strategies using MIRI/LRS eclipses and phase curves to detect signs of atmospheric heat redistribution on rocky planets, and concluded that eclipses were a more efficient way to do this, primarily using the results of \citet{mansfield2019identifying} to discard the possibility of false positives caused by high-albedo bare-rock planets. 

\citet{lustig2019detectability} simulated similar emission spectra for the TRAPPIST-1 system using both bare-rock and atmospheric models, demonstrating the large potential effect of atmospheric CO$_{2}$ absorption on the observed emission around 15$\mu$m. \citet{ih2023constraining} analysed the observed emission from TRAPPIST-1 b in the MIRI F1500W filter using similar models, showing the possibility of deriving additional information by combining it with observations in the F1280W filter.

This previous work has been used to interpret observations of rocky exoplanets, such as the secondary eclipse depths of TRAPPIST-1 b and c with the F1500W filter presented in \citet{greene2023thermal} and \citet{zieba2023no}. The observation of TRAPPIST-1 b in \citet{greene2023thermal} was consistent with the emission from a bare-rock planet with zero Bond albedo. The emission from TRAPPIST-1 c in \citet{zieba2023no} was lower than that expected for a black body, and could be consistent with a bare-rock surface with a non-zero Bond albedo, or an atmosphere with a weak CO$_{2}$ absorption feature. This analysis in \citet{zieba2023no} demonstrated the degeneracy inherent in interpreting observations with a single photometric point. Day-side emission in a single bandpass can be modified by the planetary Bond albedo or atmospheric heat redistribution, or by surface and atmospheric spectral features.  Other eclipse observations of rocky planets with JWST have also shown day-side emission consistent with a black body \citep{mansfield2024no,xue2024jwst}.

\citet{ducrot2023combined} used observations on TRAPPIST-1 b in two photometric filters to break some degeneracies inherent in single-filter observations. They combined F1500W and F1280W observations to reveal weaker emission in the F1280W bandpass than the F1500W bandpass published previously in \citet{greene2023thermal}. In the suite of models fitted, this combined dataset was consistent with a high-albedo surface or a pure-CO$_{2}$ atmosphere with an inversion.

\citet{kreidberg2019lhs} showed an alternative approach, measuring an entire phase curve of thermal emission in the 4.5 $\mu$m bandpass of the Spitzer Space Telescope over the orbit of the rocky planet LHS 3844 b. This measurement found a day-side emission consistent with a Bond albedo less than 0.2, but also a night-side emission consistent with zero. This was evidence supporting the lack of a substantial atmosphere, which would have transported heat to the night-side and likely produced a non-zero emission there (although degeneracies with an atmospheric radiating level and planetary albedo still apply). This type of method was previously demonstrated by \citet{seager2009method}, which laid out the possibility of detecting an atmosphere by finding deviations from the thermal phase curve expected for a bare-rock planet.

\citet{demory201655cnce} measured a similar thermal phase curve with the same instrument for the ``lava planet'' 55 Cancri e, and detected non-zero night-side emission that was suggested to be consistent with an atmosphere transporting heat to the night-side, although a conservative reanalysis suggested that zero night-side emission could not be ruled out \citep{mercier2022revisiting}. \citet{zhang2024gj} measured a phase curve of the rocky planet GJ 367b with the MIRI LRS instrument on JWST. Both the day-side and night-side emission from this planet were consistent with the emission expected from a bare-rock with zero albedo, with no heat redistributed to the night-side. These observations show the opportunity for phase curves to break the degeneracy inherent in the day-side emission.

In this study, we simulate secondary eclipse observations and phase curve observations for a suite of bare-rock and atmospheric models, to explore their degeneracies and to suggest optimal observing strategies for unambiguous detections of atmospheres. We describe our bare-rock model and atmospheric model in Section \ref{sec:methods}, and then present simulated observations from both in Section \ref{sec:results}. We discuss how our results can inform future observing strategies and analysis in Sections \ref{sec:discussion} and \ref{sec:conclusions}, and explore the differences in our methodologies that lead to different conclusions from \citet{mansfield2019identifying} and \citet{koll2019identifying}, as we favour phase curve observations over eclipse observations.

\section{Methods}\label{sec:methods}

\subsection{Converting Reflectance Data to Planetary Albedo}

We use reflectance data from the RELAB database \citep{milliken2021relab}, and follow the process described in \citet{hu2012theoretical} and \citet{hapke2012theory} to derive the reflectances and albedos that we need for the surface model. For each surface listed in Table \ref{tab:surfaces}, we combine a bidirectional visible reflectance spectrum with a biconical near- and mid-infrared reflectance spectrum, scaling the latter to align with the former following the RELAB manual \citep{milliken2021relab} (except for the ``Frankenspectra'' data which have already been combined). All the bidirectional reflectance data were measured with an incidence angle $i=30^{\circ}$ and an emission angle $e=0^{\circ}$, giving a phase angle $g=30^{\circ}$. RELAB provides the wavelength-dependent reflectance factor denoted $\mathrm{REFF}$ in \citet{hapke2012theory}, which we convert to the bidirectional reflectance $r_{\mathrm{dd}}(\lambda, i, e, g)$:

\begin{table*}
    \centering
    \begin{tabular}{r|p{3.0cm}|c|c|c}
       \textbf{Surface} & \textbf{RELAB Data} & \textbf{Grain Size} & \textbf{Petrogenesis} & $\mathbf{A_{B}}$\\
       \hline
       Tholeiitic basalt & RB-CMP-037, c1rb37, bir1rb037 & slab & volcanic igneous rock & 0.34 \\
       Alkaline basalt (large) & AN-G1M-008-C, c2an08c, bir1an008c & 80--160 $\mu$m & volcanic igneous rock & 0.45 \\
       Alkaline basalt (small) & AN-G1M-008-B, c1an08b, bir1an008b & 36--80 $\mu$m & volcanic igneous rock & 0.77\\ 
       Trachybasalt & AN-G1M-005-B, c1an05b, bir1an005b & 36--80 $\mu$m & volcanic igneous rock & 0.37 \\ 
       Tephrite & AN-G1M-006-B, c1an06b, bir1an006b & 36--80 $\mu$m & volcanic igneous rock & 0.53 \\ 
       Andesite & AD-REA-003-W, cwad03, n1ad03w & slab & volcanic igneous rock & 0.40 \\
       Phonolite & AN-G1M-012-B, c1an12b, bir1an012b & 36--80 $\mu$m & volcanic igneous rock & 0.53 \\ 
       Trachyte & AN-G1M-018-B, c1an18b, bir1an018b & 36--80 $\mu$m & volcanic igneous rock & 0.47 \\
       Rhyolite & AN-G1M-010-A, c1an10a, bir1an010a & \textless63 $\mu$m & volcanic igneous rock & 0.71 \\ 

        \hline
       
       Gabbro & HK-H1T-001, c1hk01, bir1hk001 & \textless2000 $\mu$m & plutonic igneous rock & 0.60 \\
       Norite & AN-G1M-019-B, c1an19b, bir1an019b & 36--80 $\mu$m & plutonic igneous rock & 0.45 \\ 
       Diorite & AN-G1M-017-A, c1an17a, bir1an017a & \textless35 $\mu$m & plutonic igneous rock & 0.68 \\ 
       Granite & AN-G1M-011-A, c1an11a, bir1an011a & \textless63 $\mu$m & plutonic igneous rock & 0.67 \\ 

        Harzburgite & FB-JFM-040-P, cpfb40, nafb40p & 25--500 $\mu$m & mantle residue & 0.69 \\ 
        Lherzolite & FB-JFM-008-P, cpfb08, nafb8p & 25--500 $\mu$m & fertile mantle & 0.58 \\ 
       
       Basalt glass & BE-JFM-060, c1be60, bir1be060 & \textless25 $\mu$m & volcanic igneous rock & 0.62 \\ 
        Basalt tuff & BU-WHF-030, c1bu30, bir1bu030 & \textless400 $\mu$m & pyroclastic ejecta & 0.57 \\
        
       \hline
       Lunar mare basalt & LS-CMP-001, n1ls01, s1ls01 & \textless500 $\mu$m & volcanic igneous rock &  0.34\\
       Lunar anorthosite & LR-CMP-224, c1lr224, bir1lr224 & \textless125 $\mu$m & primitive flotation crust & 0.82 \\  
       Basaltic shergottite & DD-MDD-028, c1dd28, bir2dd028 & \textless50 $\mu$m & igneous rock & 0.50 \\
       Mars breccia & MT-JFM-263, camt263, biramt263 & slab & impact breccia\textsuperscript{\textdagger} & 0.38 \\
       
       Albite (dust) & albite$\_$ALI \newline & \textless 1.6 $\mu$m & -  & 0.90 \\
       Magnesium sulfate & CC-JFM-019, c1cc19, s1cc19 & \textless 25 $\mu$m & - & 0.81 \\
       Pyrite & pyrite\_SA-25G \newline & \textless0.03 $\mu$m & - & 0.34 \\       
       Hematite & hematite\_SA-500G \newline & \textless0.03 $\mu$m & - & 0.55\\ 
       
    \end{tabular}

    \caption{The surface types used in the bare-rock model described in Section \ref{subsec:bare_rock_model}. The single-scattering albedo $w$ for each surface, and a script to convert it to different albedos, is available at \url{https://doi.org/10.5281/zenodo.13691959}. The horizontal lines denote the conceptual groups of ``Extrusive Igneous Rocks'', ``Other Terrestrial Igneous Rocks'', and ``Extraterrestrial Rocks and Other Minerals'' in which these surfaces are plotted in Figure \ref{fig:all_albedo_spectra}. The ``RELAB Data'' column lists the identifiers for the sample datasets in the RELAB database. The Bond albedo values $A_{B}$ are calculated for the stellar spectrum of TRAPPIST-1. \footnotesize{\textsuperscript{\textdagger}\citet{lagain2022early}.}}
    \label{tab:surfaces}
\end{table*}

\begin{equation}
     r_{\mathrm{dd}}(\lambda, i, e, g) = \frac{\mu_{0}}{\pi} \mathrm{REFF}(\lambda, i, e, g)
\end{equation}

where $\mu_{0}=\cos(i)$. We then convert $r_{\mathrm{dd}}(\lambda,i,e,g)$ to the single scattering albedo $w(\lambda)$ following \citet{hapke2012theory}\footnote{Note that like \citet{lyu2024super} we omit the opposition effect, which is included by \citet{hu2012theoretical} in a similar derivation.}:

\begin{equation}
    r_{\mathrm{dd}}(\lambda, i, e, g)=\frac{w(\lambda)}{4 \pi} \frac{\mu_0}{\mu_0+\mu} H\left(\mu_0\right) H(\mu),
\end{equation}

where $\mu = \cos(e)$. We use the two-stream approximation for the $H$ functions as described in \citet{hapke2012theory}:

\begin{equation}
    H(x) \simeq \frac{1+2 x}{1+2 \gamma(\lambda) x},
\end{equation}

where $\gamma(\lambda) = \sqrt{1-w(\lambda)}$. We then convert the derived $w(\lambda)$ to the directional-hemispheric reflectance $r_{\mathrm{dh}}(\lambda)$:

\begin{equation}
    r_{\mathrm{dh}}(\lambda) = \frac{1-\gamma(\lambda)}{1+2\gamma(\lambda)\mu_{0}},
\end{equation}

where $\mu_{0}$ is the incidence angle for the ``direction'' in the directional-hemispheric reflectance. In our bare-rock model this is the local solar zenith angle; $r_{\mathrm{dh}}(\lambda)$ then tells us the fraction of incoming stellar radiation scattered in all directions by the surface.

We also derive the spherical reflectance $r_{s}(\lambda)$, again following \citet{hapke2012theory} and using the approximation:

\begin{equation}
    r_{s} \simeq \frac{1-\gamma(\lambda)}{1+\gamma(\lambda)}\left(1-\frac{1}{3} \frac{\gamma(\lambda)}{1+\gamma(\lambda)}\right),
\end{equation}

which then defines the emissivity of the surface as $\epsilon(\lambda) = 1 - r_{s}(\lambda)$. Figure \ref{fig:all_albedo_spectra} shows the spherical reflectance $r_{s}(\lambda)$ for all our modelled surfaces. Several simplifying assumptions have been made in deriving the spherical and directional-hemispheric reflectances from the raw reflectance data from the RELAB database, so these quantities may be different for real planetary surfaces.

The spherical reflectance $r_{s}(\lambda)$ can be thought of as the wavelength-dependent equivalent of the Bond albedo. Table \ref{tab:surfaces} lists the Bond albedo for each surface, which is the fraction of incident irradiance scattered in all directions. We derive this by integrating the incoming stellar spectrum on the planetary day-side, multiplied by the directional-hemispheric reflectance $r_{dh}(\lambda)$, over the day-side and over all wavelengths. We use TRAPPIST-1 to calculate the Bond albedo values in Table \ref{tab:surfaces}. The values will vary depending on the stellar spectrum, which is taken into account when simulating the other planets in our sample. The Bond albedos in Table \ref{tab:surfaces} are generally higher than the fractional decrease in F1500W emission seen in Figure \ref{fig:compare_emission}, because the surfaces generally have higher emissivity at longer wavelengths \citep{mansfield2019identifying}. 



\subsection{Bare-Rock Planet Model}\label{subsec:bare_rock_model}

We use a numerical model of a bare-rock planet with a wavelength-dependent albedo to simulate the surface temperature and resulting thermal emission from our list of targets. We define a 10x10 grid of surface points on the day-side, and calculate their temperatures by balancing the total downward flux $F_{d}$ and the upward flux $F_{u}$ at each grid point on the surface:

\begin{equation}
    F_{d}(\lambda) = \int [ 1-r_{\mathrm{dh}}(\lambda) ]F_{S}(\lambda)\ d\lambda,
\end{equation}

\begin{equation}
    F_{u}(\lambda) = \int \epsilon(\lambda) B(\lambda, T_{\mathrm{surf}})\ d\lambda,
\end{equation}

where $F_{S}(\lambda)$ is the incoming stellar spectrum, and $B(\lambda,T)$ is the Planck function for temperature $T$. These are integrated over all modelled wavelengths and balanced, giving $\int F_{d}(\lambda) d\lambda = \int F_{u}(\lambda) d\lambda $ at each grid point. We use the stellar spectrum of the member of the SPHINX model grid with the most similar parameters to the modelled star \citep{iyer2023sphinx}, scaled to match the instellation of the effective temperature of the modelled star. We numerically balance the upward and downward fluxes to solve for the local surface temperature $T_{\mathrm{surf}}$ at each day-side grid point for a particular surface type, planet, star, and solar zenith angle. Finally, we integrate the emitted flux seen by an observer from each point on the 10x10 planetary surface grid to derive the planetary spectrum.

\subsection{Planetary Atmosphere Model}\label{subsec:atmos_model}

We use AGNI\footnote{\url{https://github.com/nichollsh/AGNI}} \citep{Nicholls2025AGNI}, a 1D radiative-convective atmosphere model, to simulate the thermal emission from a range of atmospheres on our list of targets. The radiative transfer is implemented under the two-stream and correlated-k approximations through the widely used model SOCRATES \citep{edwards_studies_1996,manners2017socrates,amundsen2017socrates}. We use gas opacity data from the DACE database\footnote{\url{https://dace.unige.ch/opacityDatabase}} \citep{grimm_database_2021} for the H$_{2}$O and CO$_{2}$ atmospheres we simulate here (shown in Figure \ref{fig:all_albedo_spectra}). Collision-induced absorption and Rayleigh scattering are included. As above, we use the SPHINX grid of stellar spectra for the incoming stellar flux \citep{iyer2023sphinx}. AGNI couples the radiative transfer to a convective model using mixing length theory \citep{joyce_mlt_2023,lincowski_trappist_2018}. The temperature structure of the atmosphere is found by solving for the state that conserves energy fluxes through each model level, subject to the boundary conditions. The effective temperature is set to zero, meaning that the planet is in radiative equilibrium with the star and is not undergoing secular cooling. A full description of the methods in AGNI can be found in \citet{Nicholls2025AGNI}. 

Table \ref{tab:atmospheres} shows the suite of atmospheric models that we analyse in this paper. The heat redistribution factor scales the instellation applied to the model, and is a simple representation of heat redistribution from the day-side to the night-side by an atmosphere. This is a highly simplified representation of the real process of dynamical heat transport to the night-side, which would be better modelled by parameterised advection \citet{lincowski2023t1c}. As the purpose of the redistribution in our model is just to produce a non-zero brightness temperature on the night-side, we assess that our simple method is sufficient for our purposes and leave more realistic modelling to future work.

\begin{table*}
    \centering
    \begin{tabular}{c|c|c}
       Surface pressure (bar) & Composition & Heat redistribution factor \\
       \hline
       1 &  N$_{2}$ + $10^{0}$ ppm CO$_{2}$ & [1, 0], [0.5, 0.5] \\
       1  & N$_{2}$ + $10^{3}$ ppm CO$_{2}$ &  [1, 0], [0.5, 0.5], [0.75, 0.25]\\
       1 &  N$_{2}$ + $10^{6}$ ppm CO$_{2}$ & [1, 0], [0.5, 0.5]\\
       1 &  N$_{2}$ + $10^{0}$ ppm H$_{2}$O &  [1, 0], [0.5, 0.5]\\
       1  & N$_{2}$ + $10^{3}$ ppm H$_{2}$O &  [1, 0], [0.5, 0.5], [0.75, 0.25]\\
       1 &  N$_{2}$ + $10^{6}$ ppm H$_{2}$O & [1, 0], [0.5, 0.5]\\
       10 &  N$_{2}$ + $10^{0}$ ppm CO$_{2}$ &  [1, 0], [0.5, 0.5]\\
       10  & N$_{2}$ + $10^{3}$ ppm CO$_{2}$ &  [1, 0], [0.5, 0.5], [0.75, 0.25]\\
       10 &  N$_{2}$ + $10^{6}$ ppm CO$_{2}$ & [1, 0], [0.5, 0.5]\\
       10 &  N$_{2}$ + $10^{0}$ ppm H$_{2}$O &  [1, 0], [0.5, 0.5]\\
       10  & N$_{2}$ + $10^{3}$ ppm H$_{2}$O &  [1, 0], [0.5, 0.5], [0.75, 0.25]\\
       10 &  N$_{2}$ + $10^{6}$ ppm H$_{2}$O & [1, 0], [0.5, 0.5]
    \end{tabular}
    \caption{The atmospheric properties used in the suite of AGNI atmospheric models described in Section \ref{subsec:atmos_model}. The bracketed pair of values defining the modelled heat redistribution determines the fraction of the incidence irradiation assigned to each of the two columns defining the day-side and night-side. Therefore, [1, 0] corresponds to no heat redistribution, [0.5, 0.5] to full redistribution, and [0.75, 0.25] to the partial redistribution used to calculate the phase curves in Section \ref{subsec:f1500w_phase_curve}.}
    \label{tab:atmospheres}
\end{table*}

\subsection{Simulating MIRI Emission Observations}

We simulate a collection of planets based on the preliminary ``Targets Under Consideration'' (TUC) list\footnote{\url{https://outerspace.stsci.edu/pages/viewpage.action?pageId=257035126}} planned for the survey programme proposed by \citet{redfield2024report}. We added TRAPPIST-1 b and TRAPPIST-1 c to this list for comparison with existing observations, ranked the list by MIRI F1500W SNR, and retained the top 30 targets by this metric. The omitted targets span a similar range of equilibrium temperatures as the top 30 targets, but have greater observational uncertainty.

We convert the planetary spectra of the bare-rock and atmosphere models to observations in the MIRI 1500W and 1280W filters, following the method in \citet{luger2017planet}. We use the estimated eclipse SNR from \citet{luger2017planet} to estimate the uncertainty of eclipse measurements (so our plots show $1\sigma$ error bars). We calculate later the uncertainty on each point in a simulated phase curve observed with either of these filters, following the method of \citet{luger2017planet}\footnote{\url{https://github.com/rodluger/planetplanet}}. We found that these simple estimates underestimated the uncertainty on real measurements \citep[e.g.,][]{greene2023thermal,zieba2023no}, so we conservatively scaled up our estimated errors on both the simulated eclipses and phase curves by $50\%$ to match the real uncertainty.

We simulate phase curves by modelling the day-side and night-side emission due to two separate 1D atmospheric models, with $25\%$ of the day-side heating redistributed to the night-side. We define the heat redistribution in this way \citep[rather than the $f$ parameter defined in][]{hansen2008absorption}, in order to refer directly to the heating applied to each of the 1D models. We set the resulting emission from the day-side and night-side columns as the maximum and minimum of a sinusoidal phase curve with its peak at zero phase (secondary eclipse). We do not add a phase offset to this simulated phase curve, despite the prevalence of such offsets for gaseous tidally locked exoplanets \citep{parmentier2017handbook}. An offset due to atmospheric dynamics would provide very strong evidence for the presence of an atmosphere, but many 3D simulations of rocky exoplanet atmospheres of this type do not predict significant phase curve offsets \citep{kane2021phase,hammond2021rotational, turbet2022trappist}. We therefore note that a phase curve offset is another possible source of evidence for an atmosphere, but we do not rely on it for our conclusions.




\subsection{Choice of surfaces}

We consider ``bare-rock'' endmember materials which are conservatively plausible from a geological perspective, as well as those which, despite not likely dominating most planetary surfaces, result in high $A_{B}$ and would therefore be the most dangerous false positives for atmospheres. The surfaces of real planets, including relatively geologically-inactive ones, are macroscopic mixtures---even Mercury has volatile-rich deposits at its poles \citep[and references therein]{rodriguez_mercury_2023}---nevertheless, mixing between endmembers would not affect the spread in bare-rock $A_{B}$, being what we are interested in. With the exceptions of MgSO$_4$ (magnesium sulfate), FeS$_2$ (pyrite), and Fe$_2$O$_3$ (hematite), we focus on natural rather than synthetic samples to capture mineralogical variability in the field (e.g., non-stoichiometric components). This list, in Table \ref{tab:surfaces}, is not exhaustive for two reasons: first, models cannot predict the detailed surface geology of an exoplanet from observable bulk properties, so our best efforts can only approximate a plausible spread; second, we found that certain surface compositions which may be thermodynamically and geologically plausible could not be included due to data availability constraints. 

The data availability is primarily limited by \textit{(i)} which samples are available on the RELAB database \citep{milliken2021relab}, and \textit{(ii)} the necessity for a sample to have both bidirectional reflectance spectra in the visible and near-infrared, as well as biconical reflectance spectra at longer wavelengths. Both datasets are needed to cover the range of stellar spectra and planetary thermal emission that we model. We accounted for grain size variations, as described below, but could not include the effects of surface temperature on observed reflectance spectra \citep[which may be non-monotonic;][]{bott_effects_2023}. 

Partial melting of a silicate mantle produces an igneous crust. Over most of modern Earth (its oceanic crust), this rock takes the particular form of mid-ocean ridge basalt. If many rocky exoplanets experience or have experienced mantle partial melting as a way to lose heat during mantle convection \citep[e.g.,][]{kite2009geodynamics}, then analogous rock types may be expected in high abundance on bare-rock planet surfaces. However, the bulk silicate compositions of exoplanets need not be the same as rocky planets in the solar system, which in itself would lead to a diversity in oceanic crust-analogues, even before divergent thermal histories of planets are invoked \citep{guimond2024stars}. To capture this diversity, we include a wide variety of terrestrial igneous rock samples from alkaline and sub-alkaline magma series (the well-known progressions from mafic to felsic compositions upon melting), many being reported in \citet{nair2017geochemical} and with detailed geologic settings given there; as well as ultramafic samples with available measurements (lherzolite, harzburgite). These samples thus qualitatively enact some of the expected variability in exoplanet mantle melt fractions, oxygen fugacities, and bulk refractory compositions. We do not exclude felsic volcanic rocks (e.g., rhyolite, phonolite) from this list---although Earth's granitic continental crust formed in the presence of liquid water, anhydrous fractional melting can also produce felsic rocks and not necessarily in very low volumes \citep{shellnutt_petrological_2013}.

We also consider materials that appear less abundantly across the planetary surfaces of the inner solar system, yet remain feasible for unknown exoplanets. In particular, the loss of an early atmosphere---or prolonged but weak outgassing \citep[see][]{foley_exoplanet_2024}---may remain imprinted in surface mineralogy, via high-temperature gas-rock reactions that proceed geologically quickly and in the absence of water \citep[e.g.,][]{zolotov_solid_2007, cutler2020experimental, filiberto_presentday_2020, teffeteller2022experimental}. Sulfate minerals could be produced by reactions between Mg-silicates and SO$_2$ gas, demonstrated experimentally and proposed for Venus and other planets \citep[e.g.,][]{renggli2019implications, berger2019experimental, rimmer2021hydroxide, byrne2024atmospheres}. A variety of chemical weathering paths produce hematite \citep[e.g.,][]{fegley1995basalt}. 
Hematite may also result from escape of an early envelope---oxygen left behind would oxidise FeO in the silicate, as estimated in \citet{kite_water_2021}. 
Conversely, enough H$_2$ in the early envelope (with respect to the availability of O) could instead imprint reducing conditions upon the surface, stabilising metals and sulfides \citep{kite_water_2021, schlichting_chemical_2021}. Such metals and sulfides are also common in enstatite meteorites in our solar system. Planets with reducing surfaces may additionally form through incomplete core-mantle segregation during magma ocean crystallisation \citep{Lichtenberg21} or accretion of sulfide-rich building blocks in certain systems \citep{jorge2022forming}.

Lastly, we include rock samples of extraterrestrial origin and of other genetic backgrounds; namely, lunar samples, martian meteorites, quickly-cooled basaltic glass, and samples representing ejecta. 
Again, these samples can represent only a slice of possibilities for real planet surfaces. Volcanic ash, for instance, can exhibit very high albedo \citep{jones_climatic_2007}, so its possible presence on a particularly volcanic planet presents a large false positive risk for an atmosphere (E. S. Kite, personal communication). We include one sample of basalt tuff (lithified volcanic ash), but finer-grained and differently-composed samples were unavailable. A famous example of a high albedo surface which we do include is lunar anorthosite rock, in this case formed by its staying afloat upon a crystallising magma ocean. Such primitive flotation crusts of lower-density, bright felsic rocks are only observed in the modern solar system on sub-planet-size bodies \citep[see][]{frossard_evidence_2019}, but we do not strictly rule out the possibility on TUC planets, which has not strictly been ruled out for primordial Earth \citep{harrison_hadean_2009}. Not all flotation crusts are so reflective, however---graphite is another material of low-enough density to float on a magma ocean, and its presence in Mercury's crust at weight-percent levels may be part of the cause for low albedos there \citep{peplowski_remote_2016, keppler_graphite_2019}.


Micrometeorite impacts on silicate crust produce small amounts of iron metal, or, again, graphite, and in this way decrease the single scattering albedo of airless surfaces in the solar system and presumably outside of it \citep{cassidy_effects_1975, lyu2024super}. However, space weathering, which groups this and multiple other processes, remains not well understood outside of the Moon and affects different materials differently \citep[e.g.,][]{gaffey_space_2010, moroz_space_2014, domingue2014mercury, pieters_space_2016, dukes_space_2016, kaluna2017simulated, yumoto2024comparison}; space weathering's effects on albedo should be investigated in future work systematically upon acquiring the necessary data. On real planets, weathering competes with resurfacing processes to determine the ``freshness'' of the surface, which will also not be spatially homogeneous. The net effects of these processes might be ground-truthed via disk-integrated spectra of airless bodies in the solar system \citep{madden_catalog_2018}, although we were unable to use such remote sensing observations for the data availability reasons mentioned above. 

As surface albedos depend strongly on regolith grain size \citep[e.g.,][]{maturilli_komatiites_2014}, we ensured that a variety of grain sizes are included in Table \ref{tab:surfaces}, from very fine dust (\textless1 $\mu$m) to outcrop. In fact, Table \ref{tab:surfaces} indicates that grain size has as big an effect as composition in controlling $A_B$. For comparison, lunar regolith has a mean particle size of $\lesssim$10--100 $\mu$m, whilst Mercury's is finer \citep{gundlach_new_2013, domingue2016application}. Grain sizes of extrasolar bare-rocks are not known \textit{a priori}. Nevertheless, dusty regoliths are plausible across the TUC insofar as the velocity of impactors increases with increasing surface gravity, implying greater comminution and smaller particle sizes \citep{cintala1992impactinduced, gundlach_new_2013}---though such impacts could also affect albedo in the other direction if they generate more nanophase metals and glass. 
Hence we include feldspar mineral dust as the highest-albedo endmember material, representing a tiny grain size and being already a light-coloured mineral. Ultimately, whilst the processes that sculpt the surfaces of airless rocky exoplanets can be informed to some degree by analogy to solar system bodies, little is understood about how these and other processes might operate on planets with higher gravity, different mineralogy, and in different radiative and dynamical environments. 

We used GGchem \citep{woitke_equilibrium_2018} to test the thermodynamic stability of each surface material (as an equilibrium condensate) under surface pressures of a generously-low 10 nbar and temperatures corresponding to the range in TUC equilibrium temperatures, for various assumptions about bulk element abundances similar to \citet{byrne2024atmospheres}. The only notable materials not stable under all temperatures were sulfates, which decompose above $\sim$650--750 K depending on bulk composition. 
No carbonates, ices, or phyllosilicates were found to be stable above $\sim$300 K at 10 nbar (i.e., most of the TUC), hence their exclusion from this particular study. Future work should test whether mineral stabilities are strongly affected by other factors in the space environment, such as proton bombardment from the stellar wind \citep[e.g.,][]{mccord_thermal_2001, dukes_space_2016}. Nevertheless, whilst we chose a very low, Io-like atmospheric pressure to be conservative, the stability of phyllosilicates shifts to higher temperatures at slightly less-conservative atmospheric pressures (e.g., they appear at 400 K at 0.01 mbar). Previous theoretical work has shown that phyllosilicates can be important even at 900 K for planetary surfaces at higher pressures than investigated here \citep{herbort_atmospheres_2020}, and that OH absorbs onto silicate grains in protoplanetary disk conditions at 700 K \citep{Thi2020}. Phyllosilicate formation does not require water ice; OH could be supplied instead from organic matter, for example \citep{hirakawa2021aqueous}. These points suggest that completely ruling out any material on unknown exoplanet surfaces will be difficult. Indeed, even the rocky-ness of the TUC is not definitely known, as more than half of the targets do not have both mass and radius measured, and eight planets out of those that do are too under-dense to be consistent with silicate-iron compositions \citep{unterborn_nominal_2023}.

\begin{figure*}
    \centering
    \includegraphics[width=\linewidth]{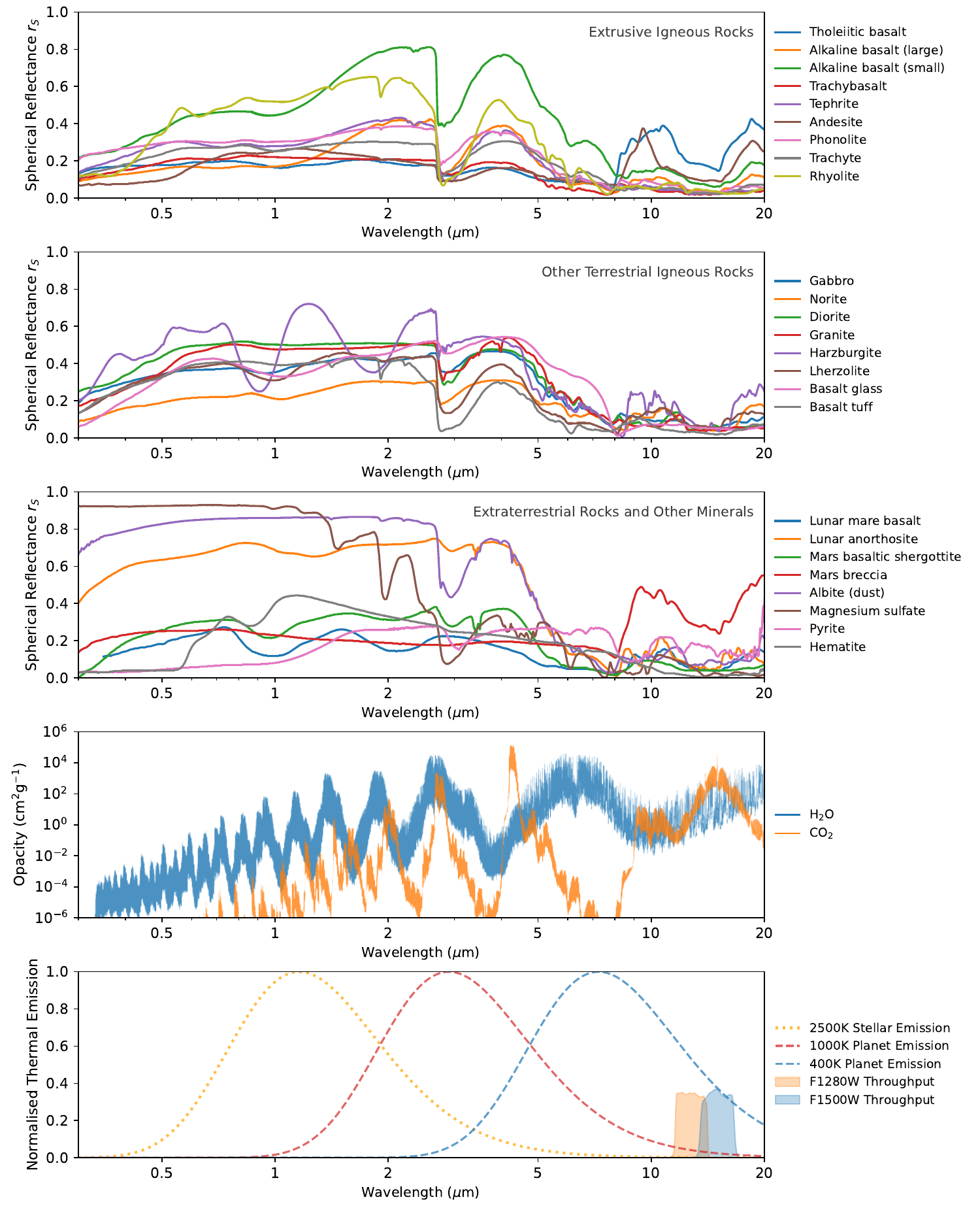}
    \caption{A summary of the input data for the bare-rock and atmospheric models, alongside the thermal emission for the typical temperatures of star and planets in this paper. The top three panels show the spherical reflectances $r_{S}(\lambda)$ for each of the planetary surfaces in Table \ref{tab:surfaces}. The single-scattering albedo $w$ for each surface, and a script to convert it to different albedos, is available at \url{https://doi.org/10.5281/zenodo.13691959}. The fourth panel shows the opacity of H$_{2}$O and CO$_{2}$ at 500K and 1 bar from DACE \citep{grimm_database_2021}. The fifth panel shows the throughputs of the F1280W and F1500W filters, as well as the black-body emission from a star at 2500 K, a planet at 1000 K, and a planet at 500K, each normalised to their maximum values}.
    \label{fig:all_albedo_spectra}
\end{figure*}

\section{Results}\label{sec:results}

\subsection{Examples of MIRI F1500W and F1280 Emission}

\begin{figure*}
    \centering
    \includegraphics[width=\linewidth]{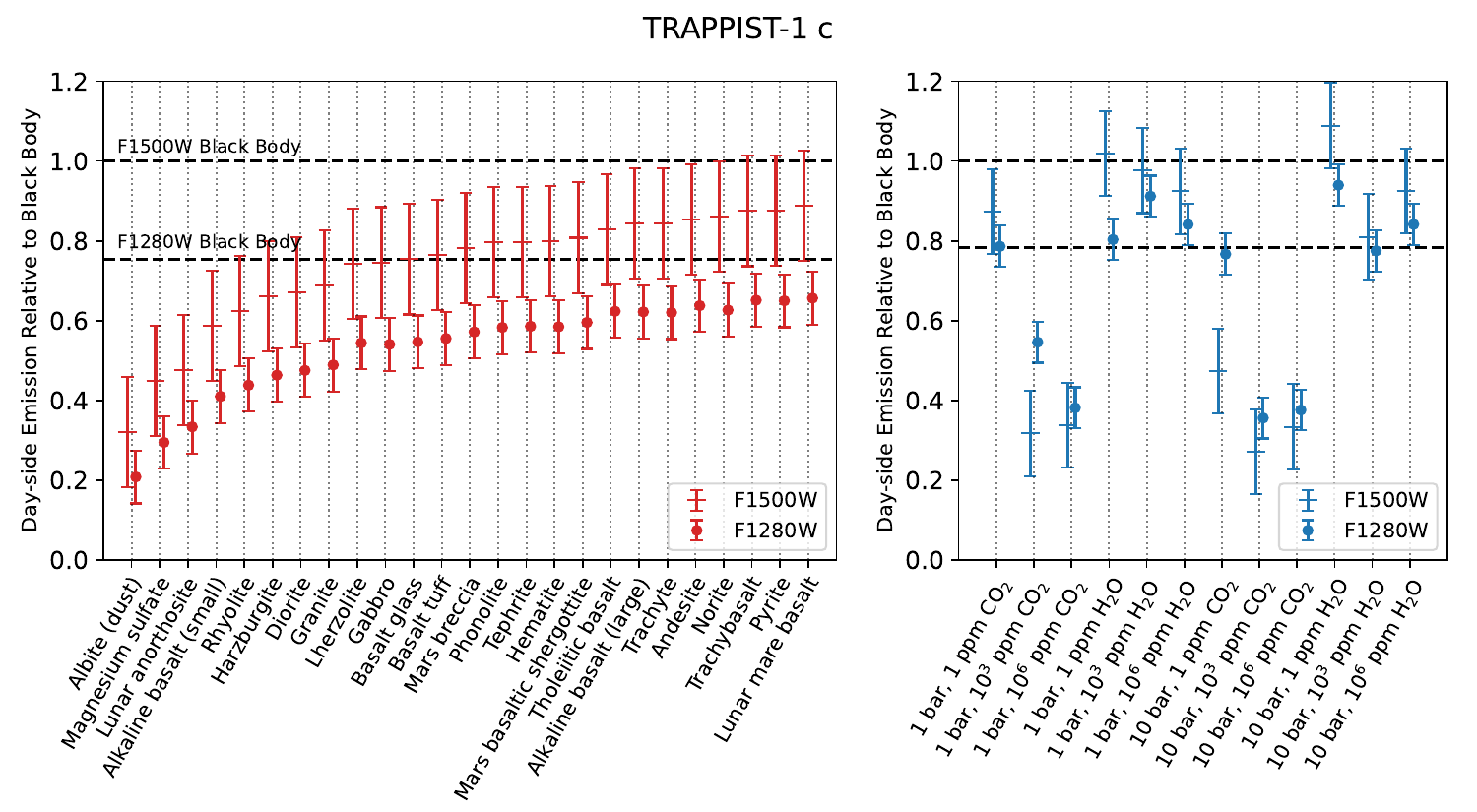}
    \includegraphics[width=\linewidth]{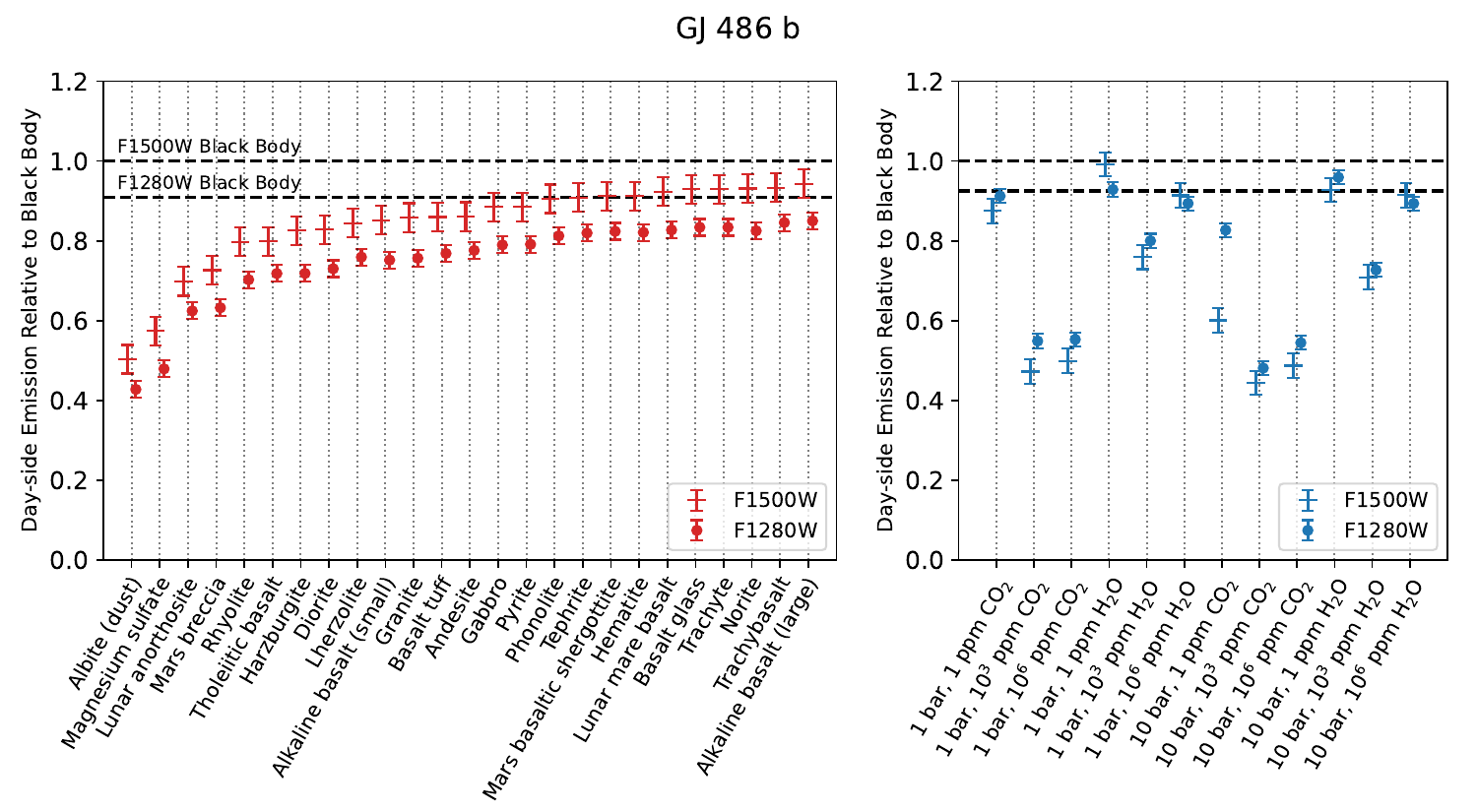}
    \caption{Simulated day-side emission  -- the average eclipse depth of five eclipses -- in the F1500W and F1280W filters for TRAPPIST-1 c and GJ 486 b, for all the surfaces in Table \ref{tab:surfaces}, and all the atmospheres in Table \ref{tab:atmospheres}. The error bars show the $1\sigma$ uncertainty on the eclipse depth, as is typically plotted for this type of measurement \citep{greene2023thermal,zieba2023no}. They are shown without heat redistribution here, which we explore in all subsequent figures. Each individual eclipse is assumed to have a baseline of four eclipse durations outside the eclipse. TRAPPIST-1 c represents the cooler planets in our sample, and GJ 486 represents the hotter planets in our sample. The average relative emission in these bandpasses is higher for GJ 486 b than for TRAPPIST-1 c, for the reasons discussed in Section \ref{subsec:f1500w_only}.}
    \label{fig:compare_emission}
\end{figure*}

Figure \ref{fig:compare_emission} shows the simulated eclipse depth and $1\sigma$ uncertainty in the MIRI F1500W and F1280W filters for TRAPPIST-1 c ($T_{\mathrm{eq}}=(341.9\pm6.6)$ K, $T_{\mathrm{eff}}=(2559\pm50)$ K, $m_{\mathrm{Ks}} =  (10.296\pm0.023)$ \citep{gillon2017seven}) and GJ 486 b ($T_{\mathrm{eq}}=(706\pm20)$ K, $T_{\mathrm{eff}}=(3291\pm75)$ K, $m_{\mathrm{Ks}} =  (6.362\pm0.018)$ \citep{caballero2022detailed}), for the suite of surface and atmosphere models in Tables \ref{tab:surfaces} and \ref{tab:atmospheres} (except the models with redistributions of $25\%$ and $75\%$). We assume that five eclipses are observed and averaged, with each eclipse observation having a baseline of four eclipse durations outside the eclipse itself (consistent with the photometric observations of TRAPPIST-1 b and c in \citet{greene2023thermal} and \citet{zieba2023no}). We highlight these two planets because JWST observations of their thermal emission have been published \citep{zieba2023no,mansfield2024no}, and they span the range of equilibrium temperatures and signal-to-noise ratios of our suite of modelled planets. The differences in the fractional emission ranges between our modelled planets are mostly due to their different signal-to-noise ratios, and their different equilibrium temperatures \citep[which scales the relative size of their emission in the F1500W and F1280W filters, for the reasons described in][]{mansfield2019identifying}. 

We simulate five eclipses because this is consistent with the five and four eclipses observed in \citet{greene2023thermal} and \citet{zieba2023no}. A uniform precision could be achieved for all the modelled planets with fewer eclipses for the targets with higher SNR, or more eclipses for the targets with lower SNR (\citet{redfield2024report} suggests a range of 2 to 15 eclipses over the proposed sample). However, this is irrelevant to our main point as we suggest that the day-side emission is highly degenerate between atmospheres and surfaces regardless of observational precision.

The emission values are normalised by the value of the emission from a black-body planet with no heat redistribution, as calculated by our bare-rock model, to highlight deviations from this value due to heat redistribution, atmospheric absorption, and surface albedo and emissivity. Figure \ref{fig:compare_emission} shows several important properties of the emission in the MIRI F1500W and F1280W bandpasses from these models. Firstly, the bare-rock emission in both cases spans a wide range of values due to the wide range of Bond albedo values shown in Figure \ref{fig:all_albedo_spectra}, overlapping with most of the atmospheric models. The emission from the atmospheric models also spans a wide range because of the variety of atmospheric structures, opacities in these bandpasses, and redistribution factors. Some of the atmospheres with H$_{2}$O have emission features in these filters, showing how a thick atmosphere can still produce emission consistent with that expected from a black body, instead of the lower emission expected due to atmospheric absorption or redistribution. This can in general be caused by greenhouse warming of the surface and subsequent observation of the surface (or near-surface) emission through a spectral window, or by the formation of a thermal inversion by strong shortwave opacity in the atmosphere. If either of these effects are sufficiently strong, the emission at a particular wavelength could be greater than the expected black-body value.

Secondly, Figure \ref{fig:compare_emission} shows how the difference in emission between the F1500W and F1280W filters is essentially the same for all the bare-rock surfaces due to their similar emissivities over this range \citep{ih2023constraining}. However, the relative emission between these filters varies for the atmospheric models, due to differences in atmospheric opacity. The largest differences are due to the significant difference in CO$_{2}$ opacity between the F1500W and F1280W bandpasses. Some (but not all) of the atmospheric models can be distinguished from bare-rock surfaces by this method \citep{ih2023constraining}.

Our models are a limited representation of all the possible surfaces and atmospheres on such a planet. Different surfaces or atmospheres could have higher albedos or emission features, or stronger differences in their F1500W and F1280W emission. Despite our wider selection of surface types than previous studies, the Bond albedos of most types of igneous extrusive rock are relatively low, resulting in relatively high day-side emission, consistent with previous work \citep{hu2012theoretical,mansfield2019identifying,lyu2024super}. The surfaces with small grain sizes, or the more unusual surfaces like magnesium sulfate, provide the highest Bond albedo values. With that said, regolith grain sizes of a few tens of microns are not unusual in the solar system and can result in surprisingly high Bond albedos even for common basalt. We do not attempt to provide a prior likelihood on the prevalence of different surface types; understanding the likelihood of high-albedo surfaces will therefore be crucial to make progress in this area.


\subsection{Detecting Atmospheres with Day-side F1500W Emission}\label{subsec:f1500w_only}

\begin{figure*}
    \centering
    \includegraphics[width=\linewidth]{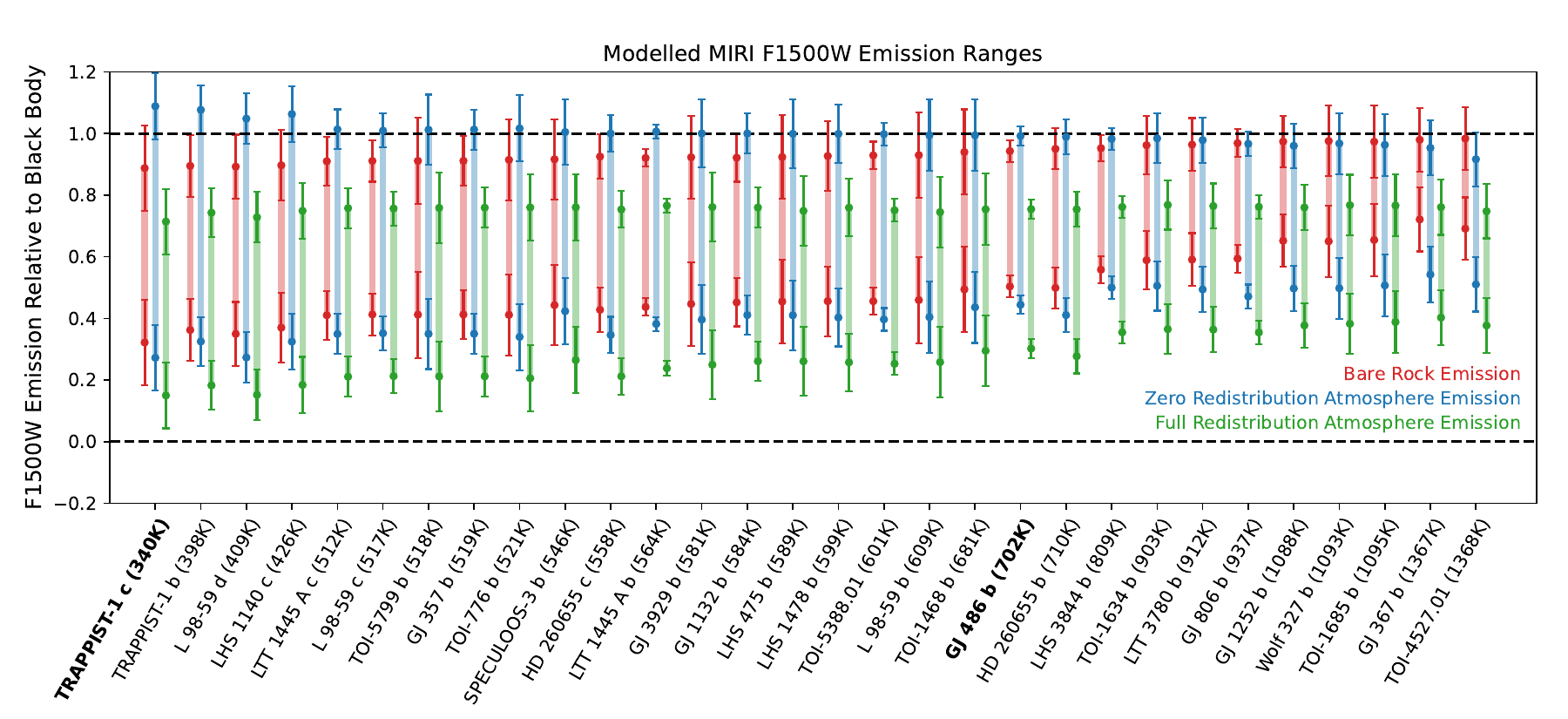}
    \caption{Comparing the ranges of possible F1500W day-side emission (for five observed eclipses) from the bare-rock model, the atmospheric model with zero heat redistribution, and the atmospheric model with full heat redistribution. Each observation is modelled as the average of five secondary eclipses, with a baseline of four eclipse durations outside the eclipse. Each plotted range compiles the relevant models from the suites like those in Figure \ref{fig:compare_emission}; for example, the left-most range shows the range of possible F1500W emission values for TRAPPIST-1 c due to all the different surface types in Table \ref{tab:surfaces}. The planets are ordered by equilibrium temperature. TRAPPIST-1 c and GJ 486 b are highlighted in bold, as they are plotted in more detail in Figure \ref{fig:compare_emission}. The bare-rock surfaces can only be distinguished from the atmospheres with zero heat redistribution on GJ 806 b and LTT 1445 A b (that is, the uncertainties on their lower limits do not overlap). The bare-rock surfaces can be distinguished from some of the atmospheres with full heat redistribution on more planets, for a subset of atmospheres with both full heat redistribution and a low CO$_{2}$ abundance.}
    \label{fig:compare_atmos_emission_names}
\end{figure*}

Figure \ref{fig:compare_atmos_emission_names} shows the range of emission values from each planet in our suite, for the bare-rock model, the atmospheric model without heat redistribution, and the atmospheric model with full global heat redistribution. For example, the left-most points show the highest and lowest values for the emission for the bare-rock model of TRAPPIST-1 c. They are joined with a shaded line to show that the models of emission span the region between these two extremes; most of the ranges are covered continuously by the variety of models in each category.

This plot is intended to show how an observation of the F1500W emission can be consistent with several potential surfaces or atmospheres. The possibility of bare-rock surfaces with high albedos \citep[see also the ``feldspathic'' and ``granitoid'' surfaces modelled in][]{zieba2023no} means that it is very difficult to conclude that a low day-side emission in the F1500W is unambiguously due to an atmosphere. Figure \ref{fig:compare_atmos_emission_names} shows that only two of the atmospheres we model with zero heat redistribution can be distinguished from the range of potential surfaces (comparing the lower end of the red and blue ranges for each planet). These are the atmospheres with 1 ppm CO$_{2}$ (which provides a strong absorption feature in the F1500W bandpass, without significantly heating the atmosphere) for GJ 806 b and LTT 1445 A b (which both have high signal-to-noise ratios).

Heat redistribution to the night-side lowers the emission from the atmospheric models (as would an increased atmospheric albedo due to clouds). Many of the CO$_{2}$-dominated atmospheres can be distinguished from bare-rock surfaces for the planets with the higher signal-to-noise ratios. However, many of the modelled planets are approaching the limit of a detectable eclipse at the low values of fractional emission around 0.25 and below in Figure \ref{fig:compare_atmos_emission_names}.

There are plausible reasons to expect lower albedo values for bare-rock surfaces in general. The lower end of the range of possible surface emission values is driven by high-albedo surfaces types in the ``Extraterrestrial Rocks and Other Minerals'' category in Table \ref{tab:surfaces}. For TRAPPIST-1 c, Figure \ref{fig:compare_emission} shows that the range of F1500W emission values due to the ``Extrusive Igneous Rocks'' and ``Other Terrestrial Igneous Rocks'' extends down to $\sim$0.45 rather than $\sim$0.20 for the whole sample of surfaces. 

Similarly, we do not model the effect of space weathering on the albedo and emissivity of the bare-rock surfaces. This could plausibly decrease the albedo of these surfaces, making them more readily distinguished from CO$_{2}$-dominated atmospheres \citep{hu2012theoretical}. However, we suggest this does not allow unambiguous detections of atmospheres.  As mentioned above, the extent of darkening due to nanophase metals at a given time reflects a competition between space weathering and (poorly-constrained) resurfacing processes acting against weathering. Whilst bare-rocks in the solar system frequently appear weathered and darker, these bodies are all small and relatively geologically inactive. The geophysical heat engine would persist for longer on more massive bodies \citep{stevenson_styles_2003, kite2009geodynamics}, promoting frequent reworking of the crust, irrespective of atmosphere retention. Thus, even if we might expect (for example) high-albedo lunar anorthosite to be weathered and become darker \citep{yamamoto_spaceweathered_2018}, we cannot guarantee that this is the case for an unknown planet. For the purposes of understanding the plausible range of emission, we do not model space weathering because it would simply move the F1500W emission of each surface upwards toward the black body value. As our range of surfaces already spans this range of emission, modelling space weathering would not add to the possible range of F1500W emission from our bare-rock models.

The ordering of the planets by equilibrium temperature in Figure \ref{fig:compare_atmos_emission_names} shows how the emission from bare-rock surfaces in the F1500W bandpass (as a fraction of the black body value) rises with planetary equilibrium temperature, most strongly at higher equilibrium temperatures. \citet{mansfield2019identifying} showed how this is due to the peak of the Planck function moving to shorter wavelengths for hotter planets, where the emissivity of the bare-rock surfaces is relatively lower (see Figure \ref{fig:all_albedo_spectra}), which makes the emission at 15$\mu$m relatively higher. This is a fractional effect on the change in emission from a black-body, so has a larger absolute effect on the lower end of the emission range than the higher end.

However, the minimum of the emission from the atmospheric models also rises with equilibrium temperature. This is because exactly the same process occurs for the emissivity of a CO$_{2}$-dominated atmosphere (the atmospheres with the lowest F1500W emission are all CO$_{2}$-dominated). Figure \ref{fig:all_albedo_spectra} shows how CO$_{2}$ has low opacity from 5 to 10 $\mu$m; cool planets emit freely in this range and so their emission at 15$\mu$m is relatively low. However, the peak of the Planck function for hotter planets is lower, from 2 to 5$\mu$m, where the average CO$_{2}$ opacity is higher. This decreases the relative emissivity in this region, which increases the relative emission at 15$\mu$m. Therefore, the increase in F1500W emission from bare-rock surfaces at higher equilibrium temperature that was identified in \citet{mansfield2019identifying} does not aid so much in distinguishing them from CO$_{2}$-dominated atmospheres, because exactly the same increase in F1500W emission at higher equilibrium temperatures occurs for CO$_{2}$-dominated atmospheres.

So far, we have focused on the possibility of false positive detections of atmospheres, when high-albedo surfaces are observed. It would also be possible to derive a false negative detection of an atmosphere, because the upper range of the emission from the modelled atmospheres extends up to the expected value for a black body bare-rock planet, and even above this value in many cases. Therefore, a measurement of emission that is consistent with the black body value does not necessarily demonstrate the absence of an atmosphere.

\subsection{Detecting Atmospheres with Day-side F1500W and F1280W Emission}\label{subsec:f1500w_f1280w}

\begin{figure*}
    \centering
    \includegraphics[width=\linewidth]{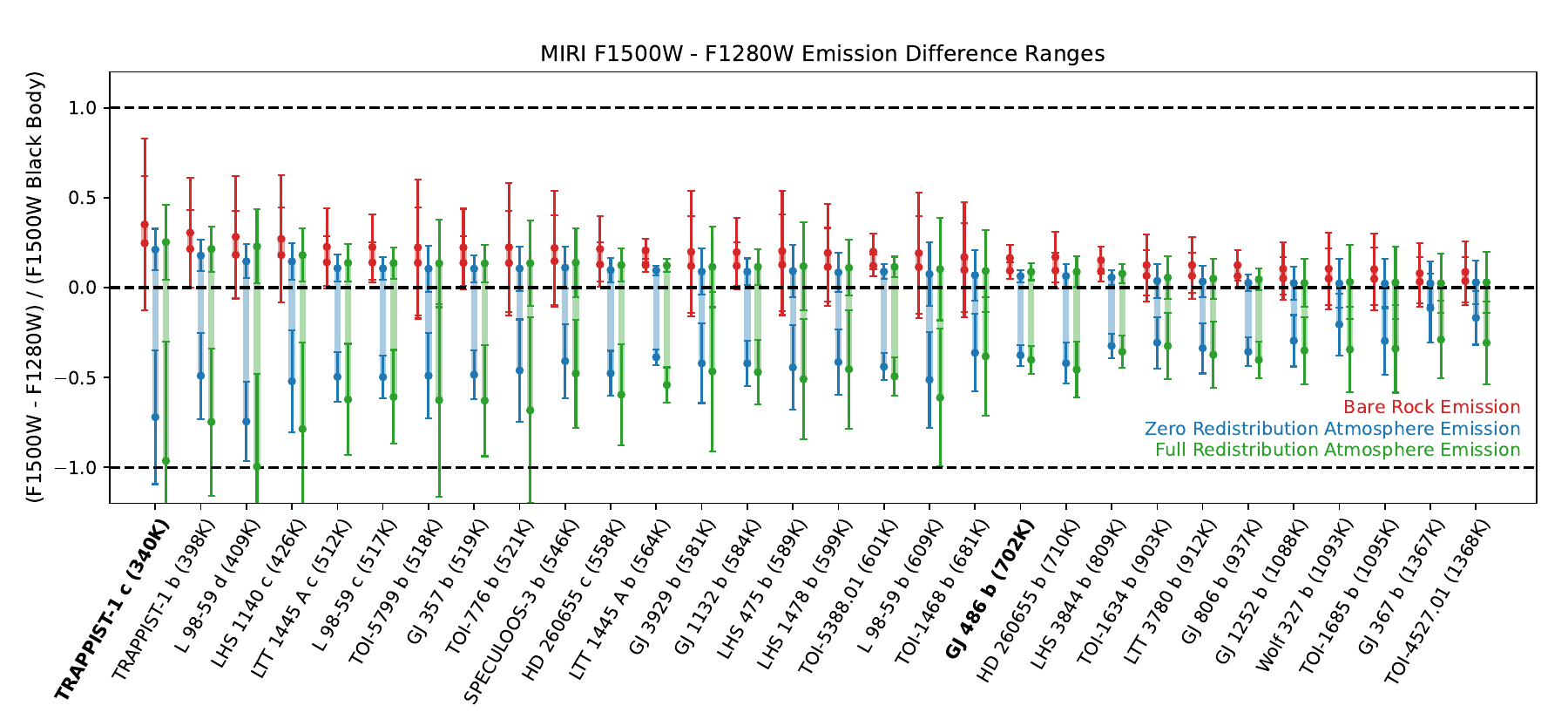}
    \caption{Comparing the ranges of the differences in day-side emission between the F1500W and F1280W filters from the bare-rock model, the atmospheric model with zero heat redistribution, and the atmospheric model with full heat redistribution. Each observation is modelled as the average of five secondary eclipses in each of the two filters, with a baseline of four eclipse durations outside the eclipse. More of the atmospheres can be detected than in Figure \ref{fig:compare_atmos_emission_names} by this metric, which makes use of the CO$_{2}$ spectral feature in the F1500W bandpass, which is weaker in the F1280W bandpass. Only the atmospheres with 1 ppm CO$_{2}$ can be reliably detected by this metric, because for higher abundances of CO$_{2}$ its opacity saturates the F1280W bandpass as well as the F1500W bandpass (see Figure \ref{fig:all_albedo_spectra}). The atmospheres with H$_{2}$O cannot be detected by this metric, because the opacity of H$_{2}$O is similar in both bandpasses (Figure \ref{fig:all_albedo_spectra}), resulting in similar emission in both filters (Figure \ref{fig:compare_emission}). The atmospheres with full heat redistribution are not much more detectable by this metric than the atmospheres with zero redistribution. This is because any redistribution reduces the emission in both the F1500W and F1280W filters by a roughly equal fraction.}
    \label{fig:compare_delta_atmos_emission_names}
\end{figure*}

Figure \ref{fig:compare_delta_atmos_emission_names} shows the difference in emission between the F1500W and F1280W filters for each of our modelled planets, normalised by the F1500W emission from a black body. This difference is almost constant for the bare-rock surfaces, but can vary for the modelled atmospheres, generally due to the CO$_{2}$ absorption feature in the F1500W bandpass. 


Our modelling in Figure \ref{fig:compare_delta_atmos_emission_names} reaches a similar conclusion for TRAPPIST-1b -- only a subset of atmospheres have a detectable difference in the relative emission in these two filters. Some other planets with a higher signal-to-noise ratio have more readily detectable differences in emission. Figure \ref{fig:compare_emission} shows that these differences are strongest for a subset of the CO$_{2}$ atmospheres (1 bar with $10^{3}$ ppm CO$_{2}$, and 1 bar with 1 ppm CO$_{2}$). For the highest concentrations of CO$_{2}$, the atmospheric opacity increases in the F1280W bandpass as well as the F1500W bandpass, so the difference in emission between them decreases. This metric does not generally detect H$_{2}$O-dominated atmospheres, as its opacity is similar in both filters (Figure \ref{fig:all_albedo_spectra}). In all cases, the results are similar whether the atmospheres have heat redistribution or not, because the emission in the bandpasses of both filters changes by a similar fraction when heat is redistributed.


Therefore, while there are some atmospheric compositions that would be detectable from the difference between these two filters, these compositions are a limited set of the total possibilities. Moreover, observing the emission in two filters requires observing twice as many eclipses as for one filter. For example, measuring five eclipses of TRAPPIST-1 b, each with a duration of the eclipse itself plus an additional out-of-eclipse baseline of four eclipse durations, takes 15.1 hours. Each separate observation also requires approximately one additional hour for telescope slewing, target acquisition, and detector stabilisation\footnote{\url{https://jwst-docs.stsci.edu/jwst-general-support/jwst-observing-overheads-and-time-accounting-overview/}}. Measuring five eclipses for two filters then takes 40.2 hours in total, which is similar to the time of 37.3 hours for a full orbital phase curve observation.

\subsection{Detecting Atmospheres with F1500W Phase Curves}\label{subsec:f1500w_phase_curve}

\begin{figure*}
    \centering
    \includegraphics[width=\linewidth]{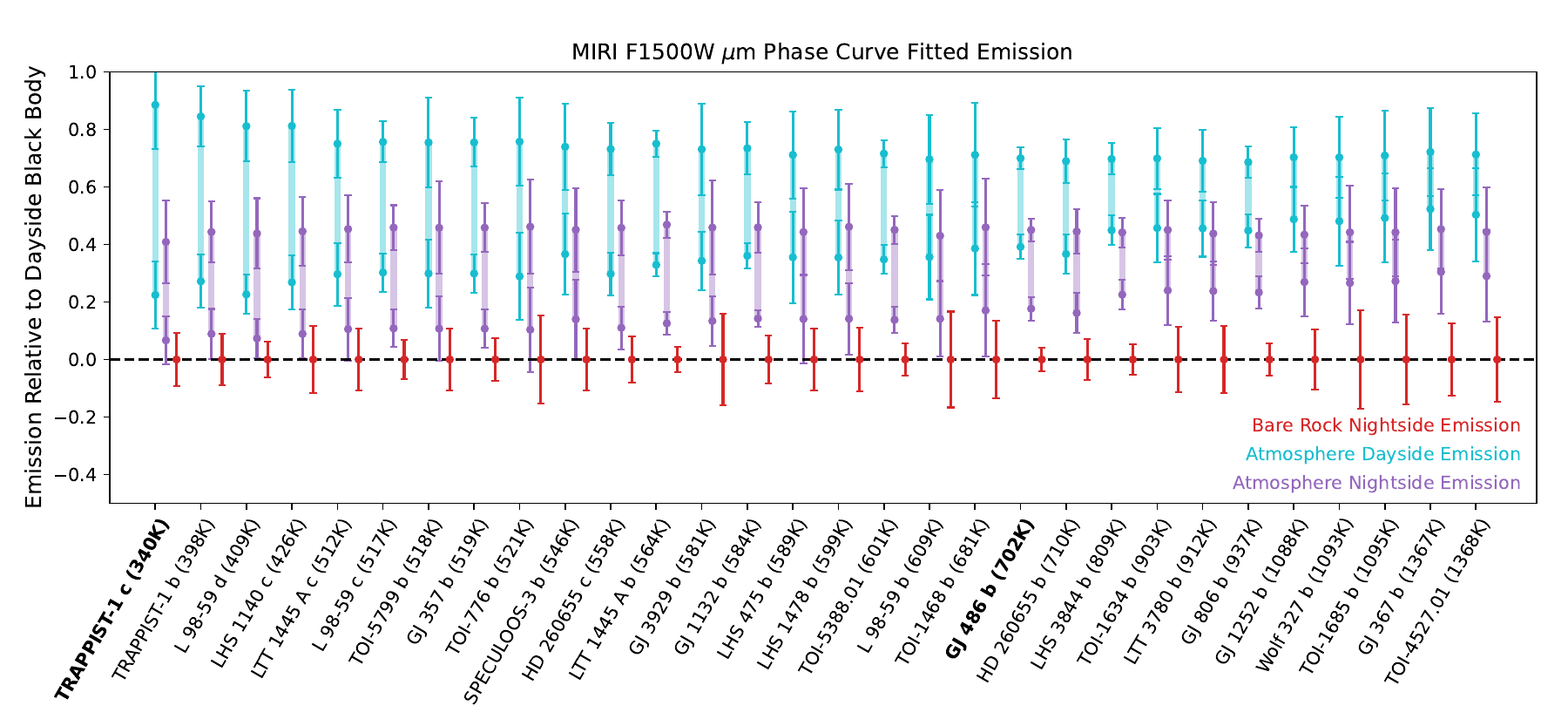}
    \caption{The ranges of modelled (and then fitted) F1500W phase curve maxima and minima, for the range of atmospheres in Table \ref{tab:surfaces}, compared to the zero flux minimum expected on the night-side of a bare-rock planet. Each observation is modelled as a phase curve over one orbit. The plotted points are the true value of the maxima and minima of the phase curves, and the error bars are the $1\sigma$ uncertainty of the values fitted with \citet{exoplanet:pymc3}. As we did for the eclipse-only observations, the error bars are conservatively scaled up by $50\%$ to ensure they are at least as large as they would be for real observations. With a heat redistribution of $25\%$ of the day-side energy to the night-side, almost every planet has detectable night-side emission. Those that do not are atmospheres with low CO$_{2}$ abundances on planets with low signal-to-noise ratios; these also have detectably weak emission from their day-side as a result. We do not include the day-side emission for bare-rock surfaces as these are shown in Figure \ref{fig:compare_atmos_emission_names} with comparable precision to that achieved by a phase curve. Moreover, our aim in the current figure is to identify unambiguous information by distinguishing non-zero night-side emission due to atmospheric heat redistribution.}
    \label{fig:compare_phase_emission_names}
\end{figure*}

This section shows simulations of the day-side and night-side emission retrieved from phase curves observed in the F1500W filter. The maximum of the phase curve is the same physical quantity as the day-side emission measured from the simulated secondary eclipses above, so is also affected by the degeneracy between surface Bond albedo, and atmospheric properties. However, the minimum of the phase curve is generally immune to these degeneracies, if it is consistent with non-zero night-side emission, as the emission from the night-side of a bare-rock planet will be negligible. A measurement consistent with zero night-side emission would be degenerate between a bare-rock surface and an atmosphere with weak heat redistribution or weak emission in the observed bandpass. Planets hotter than the sample we consider may have magma oceans whose currents could transport some heat in the absence of atmospheres, but this transport is expected to be inefficient and so would not significantly affect the surface temperature and resulting nightside thermal emission \citep{kite2016atmosphere,Meier2023, lai_ocean_2024}.

We simulate phase curves containing two eclipses, with a duration $20\%$ longer than a single orbital period. Their maximum and minimum are calculated by a 1D model for each side of the planet, with stellar heating of $75\%$ and $25\%$ of the total. We then add observational noise and fit a model of a sinusoidal phase curve with a free amplitude, mean, phase offset, and stellar baseline using \citet{exoplanet:pymc3}. These atmospheres all assume that $25\%$ of the stellar heating on the day-side is redistributed to the night-side; atmospheres with less redistribution would be less detectable by their night-side emission. This fraction is unknown for exoplanets, and depends on a range of atmospheric properties like surface pressure and composition; our choice of $25\%$ is to demonstrate that an atmosphere with enough heat redistribution can be unambiguously detected. The value of $25\%$ is consistent with the order of magnitude of the heat redistribution in the 3D atmospheric simulations and scaling theory in \citet{koll2022scaling}. It is also consistent with the outgoing longwave radiation on the night-side plotted for a variety of simulations of TRAPPIST-1 e in \citet{turbet2022trappist}. If we modelled planets with full heat redistribution (as we do for the eclipse-only observations above), their night-side emission would be even more detectable.


Figure \ref{fig:compare_phase_emission_names} shows the range of phase curve maxima (essentially the same as the day-side emission shown in Figure \ref{fig:compare_atmos_emission_names}), and the range of phase curve minima, for atmospheres 2, 5, 8, and 11 with $25\%$ heat redistribution in Table \ref{tab:atmospheres}. The emission from the ``null hypothesis'' of a bare-rock night-side is also shown for each planet, to determine which night-sides could be observationally distinguished from a bare-rock planet. Figure \ref{fig:compare_phase_emission_names} shows that almost all of these simulated atmospheres produce enough night-side emission to be distinguished from a bare-rock planet. 

A phase curve observation also provides a measurement of day-side emission as well as night-side emission, with comparable precision to that shown in Figure \ref{fig:compare_atmos_emission_names}. This information can be complementary to night-side information; for example, many of the simulated observations in Figure \ref{fig:compare_phase_emission_names} with detectable night-side emission have detectably low day-side emission in Figure \ref{fig:compare_atmos_emission_names}, providing corroborating evidence for an atmosphere.



Some atmospheres in Figure \ref{fig:compare_phase_emission_names} produce too little night-side emission to be distinguished from the bare-rock null hypothesis. These are the atmospheres with low amounts of CO$_{2}$, which emit from low pressures but have little greenhouse warming. While this may produce too little emission to be detected on the night-side, the day-side  emission will be correspondingly low (see the lowest day-side emission value for each planet in Figure \ref{fig:compare_phase_emission_names}). This provides the same evidence for atmospheric absorption on the day-side as would be provided by observing secondary eclipses only. Observing a phase curve with a minimum flux consistent with zero should not be taken as proof of the lack of an atmosphere, as thin atmospheres may redistribute too little heat. \citet{powell2024nightside} shows that optically thick night-side clouds may produce the same effect, suppressing detectable night-side emission from a thick atmosphere.

Most of the atmospheres we simulate produce night-side emission that can be distinguished from a bare-rock planet, using an observation of a single phase curve. Any atmosphere that is thick enough to redistribute a non-negligible fraction of the day-side heating to the night-side will be detectable in this way. An atmosphere detected by a phase curve will also be better characterised than one detected by secondary eclipses only, with a constraint on day-night heat transport and a potential phase curve offset.



\section{Discussion}\label{sec:discussion}

\subsection{Comparison to previous work}

We reach some different conclusions to previous studies on this topic. In this section, we compare our methodology and results to \citet{mansfield2019identifying}, \citet{whittaker2022detectability}, \citet{koll2019identifying}, and \citet{lustig2019detectability} to determine the causes of these differences.

\citet{mansfield2019identifying} presents ``a new method to detect an atmosphere on a synchronously rotating rocky exoplanet around a K/M dwarf, by using thermal emission during secondary eclipse to infer a high day-side albedo that could only be explained by bright clouds''. They used a similar methodology to \citet{hu2012theoretical} to simulate the broadband MIRI/LRS emission from eight surface types, to conclude that ``a high albedo could be unambiguously interpreted as a signal of an atmosphere for planets with substellar temperatures of $\mathrm{T}_{\mathrm{sub}}=$410–1250 K''. This significantly differs from our conclusion in Section \ref{subsec:f1500w_only}, where we suggest that high-albedo bare-rock planets could produce day-side thermal emission that is much lower than a black body planet, and therefore highly degenerate with an atmosphere. 

We have determined that the source of this discrepancy is that \citet{mansfield2019identifying} calculated the surface temperatures of each bare-rock surface using the geometric albedo values derived in \citet{hu2012theoretical} (M. Mansfield, private communication). The geometric albedo is the ratio of planetary flux at zero phase angle (from the day-side, for a tidally locked planet) to the flux from a Lambert disk \citep{seager2010exoplanet}; it is an observational quantity and does not control planetary energy balance or equilibrium temperature. The Bond albedo determines the fraction of incoming total stellar energy scattered into space in all directions, so is the quantity that controls planetary energy balance and equilibrium temperature as described in Section \ref{subsec:bare_rock_model} (note that we use its wavelength-dependent equivalent $r_{\mathrm{dh}}(\lambda)$ there).

The geometric albedo is $50\%$ lower than the Bond albedo for the surfaces modelled here and in \citet{mansfield2019identifying}, because they are assumed Lambertian \citep{seager2010exoplanet}. Using the geometric albedo therefore results in significantly higher bare-rock surface temperatures in \citet{mansfield2019identifying} than for the equivalent surfaces in our Section \ref{subsec:f1500w_only} with temperatures calculated using the Bond albedo. The difference in our methods can be seen by comparing the spherical reflectance (which gives the Bond albedo when integrated) for ``Lunar Anorthosite'' in our Figure \ref{fig:all_albedo_spectra}, to the  ``Feldspathic'' albedo value in Figure 3 of \citet{mansfield2019identifying}. Both use the same original RELAB data (LR-CMP-224), but our albedo values are approximately $50\%$ higher. This can be inspected in more detail using the datasets at \url{https://doi.org/10.5281/zenodo.13691959}.

This is why we are much more pessimistic about the prospects of detecting atmospheres with eclipses only, as our modelled bare-rock surfaces produce approximately $50\%$ weaker emission than the equivalents in \citet{mansfield2019identifying}. Another minor difference is that we consider a wider range of surface types than \citet{mansfield2019identifying}, which broadens the range of possible surface emission values, increasing the degeneracy. We note that \citet{whittaker2022detectability} applies the same methodology as \citet{mansfield2019identifying}, using the geometric albedo to calculate temperatures of bare-rock surfaces.

\citet{koll2019identifying} also tackled this issue, comparing the relative merits of phase curves, spectroscopy, and eclipse photometry, for detecting atmospheres on rocky planets. \citet{koll2019identifying} concluded that ``infrared photometry of secondary eclipses could quickly identify “candidate” atmospheres, by searching for rocky planets with atmospheres thick enough that atmospheric heat transport
noticeably reduces their day-side thermal emission compared to that of a bare-rock \textup{[\,\dots]} Candidate atmospheres can be further validated via follow-up spectroscopy or phase curves''. They favour secondary eclipses over phase curves for detecting candidate atmospheres due to the shorter observational time required. They suggest that the false positive bare-rock scenario is unlikely primarily by reference to \citet{mansfield2019identifying}, supported by discussion of low-albedo bare-rocks in the Solar System and on \citet{kreidberg2019lhs}. 

Our conclusions are similar in general to those of \citet{koll2019identifying}: we also find that eclipse photometry could identify signatures of candidate atmospheres, and that these must then be followed up with more detailed measurements. Our point of difference is that we do not discard the bare-rock false positive case following \citet{mansfield2019identifying}, because our use of the Bond albedo rather than the geometric albedo results in much lower thermal emission from bare-rock planets. We therefore suggest that one may as well proceed to the detailed phase-curve ``follow-up'' as the secondary eclipse measurements are too degenerate to contain useful information. This is ultimately a qualitative point and that the conclusion of \citet{koll2019identifying} may still apply if a strategy of initially identifying candidates for later follow-up is preferable. 


Our final comparison to previous work is with \citet{lustig2019detectability}, which focuses mainly on observational signatures of different atmospheres, but also simulated the emission signatures of bare-rock planets. Their Figure 1 shows the secondary eclipse spectrum of TRAPPIST-1 b for a variety of bare-rock surface types. As \citet{lustig2019detectability} makes clear, these spectra are calculated assuming zero Bond albedo so all have the same planetary equilibrium temperature. This results in them emitting similarly to a black body at $15\mu$m, unlike our simulations where many surfaces have much lower emission due to their high Bond albedo and lower temperatures. The main conclusion of \citet{lustig2019detectability} from these results is that ``an airless rock would likely have significantly lower spectral variation than atmospheric features''. This is consistent with our results, although we do not focus on the specifics of the spectral variation of the emission beyond the differences in the F1500W and F1280W filters in Section \ref{subsec:f1500w_f1280w}. 


\subsection{False Positives and False Negatives}

Detecting atmospheres by eclipse photometry alone depends on our understanding of the link between atmospheric thickness, global heat redistribution, and atmospheric emission in the observed filter. This link is generally suggested to be that thicker atmospheres on tidally locked planets redistribute more heat \citep{koll2022scaling}, which then emit more weakly from their day-sides. This is suggested to contrast with the higher emission from bare-rock surfaces, as igneous rocks generally have lower Bond albedos (see Table \ref{tab:surfaces}) and relatively higher emissivity in the F1280W and F1500W MIRI filters \citep{mansfield2019identifying}.

While all these statements are physically reasonable, we suggest our results demonstrate plausible false positive and false negative detections of atmospheres by eclipse observations. Firstly, the wide range of surface types we consider have a range of albedos, with the most reflective having Bond albedos over 0.5. While low-albedo igneous rocks are probably more likely surfaces \citep[especially if they are darkened by space weathering;][]{hu2012theoretical}, it is not possible to entirely rule out known or unknown high-albedo surfaces. Measuring low emission in the F1500W filter from a high-albedo bare-rock planet could therefore provide a false positive detection of an atmosphere.

Secondly, while thick atmospheres are generally likely to redistribute heat and emit more weakly in the F1500W bandpass, we suggest there are plausible effects that could counteract this. \citet{koll2022scaling} demonstrates a very compelling scaling relation between atmospheric thickness and heat redistribution in an idealised 3D atmospheric General Circulation Model (GCM), but there are many other surface and atmospheric properties that could affect heat transport such as land-mass distribution, topography, condensation, or clouds \citep{lewis2018influence,sergeev2020atmospheric}. 

Moreover, even with strong heat redistribution, an atmosphere may still emit like a bare-rock in a particular bandpass. The greenhouse effect could warm the surface beyond its equilibrium temperature, which could then emit at least as strongly as a bare-rock at wavelengths with low atmospheric opacity. This is why some of the atmospheres in Figure \ref{fig:compare_emission} emit more strongly than a black body surface would. Secondly, an atmosphere may form a thermal inversion if, for example, its visible and thermal opacities scale differently with pressure \citep{Piette2023,Zilinskas2023}. It could then produce an emission feature in the F1500W bandpass if it has high opacity in that spectral region.

Both of these effects could counteract the overall cooling effect of heat redistribution, raising the emission in a particular bandpass back up to the value expected for a black body. An observation of such an atmosphere in that bandpass could therefore provide a false negative conclusion that there is no atmosphere.  Unlike the false positives and false negatives for single-filter observations of secondary eclipses, we suggest that night-side emission can provide an unambiguous detection of an atmosphere. 

\subsection{Issues with Phase Curves}

While we suggest that phase curves can resolve the degeneracy between the day-side emission of bare-rock planets and atmospheres, there would still be challenges in making and interpreting observations of night-side emission.

Instrumental systematics are a key issue for the processing of MIRI observations, particularly a distinctive exponential ramp at the start of an observation \citep[e.g.,][]{zieba2023no,kempton2023reflective, bell2024nightside}. \citet{august2024hot} concluded that these systematics are more problematic than previously expected when analysing MIRI filter eclipse observations of rocky exoplanets. Repeated eclipse observations and phase curve observations have different advantages and disadvantages when handling these systematics. Repeated eclipse observations require a relatively simple model to be fitted, which should not be degenerate with the shape of an exponential ramp or linear trend. However, every eclipse will have its own systematic shape, requiring a new set of parameters to be fitted each time \citep{zieba2023no}. Observing a phase curve with at least two eclipses provides a single periodic measurement which can separate an exponential systematic ramp from the periodic planetary signal more effectively than separated eclipse observations \citep{hammond2024wasp43b}. However, the shape of the phase curve can be degenerate with long-period systematics, with the strongest effects on the inferred night-side emission furthest from the anchoring effect of the eclipses \citep{kempton2023reflective}. 


There is an important potential false positive for the detection of an atmosphere using a phase curve. Internal heating processes could lead to a non-zero surface geothermal heat flux at all longitudes, warming the night-side even in the absence of an atmosphere. For internal heating via the decay of radioactive isotopes in an Earth-like concentration, we expect an associated heating rate at the surface on the order of a few tens to a hundred mW\,m$^{-2}$, which is insufficient to be detected in the near- to mid-infrared spectrum with JWST \citep{Meier2021}. Tidal dissipation in the interior adds to this flux; the tidal contribution to surface heating can be estimated from the orbital parameters of the system (which would be refined with a high-precision phase curve), with some assumptions about the interior structure and rheology \citep[e.g.,][]{barr_interior_2018, hay_tides_2019, bolmont_solid_2020, 2024arXiv241207285F}. 

The observed globally-averaged surface heat flux on Io (essentially the tidal heat flux here) is a few W\,m$^{-2}$, still undetectable---though note that localised volcanoes could contribute disproportionally to the observed thermal emission. In this context it is irrelevant whether the heat generated in the interior is mostly transported at the surface by conduction through the planet's crust, like Earth, or by the advection of hot magma like Io. The planned DDT programme aims to improve constraints on the eccentricity of the target planets, which should improve the precision of the upper limit on their tidal heating.

The greatest issue with phase curves is the observational time that they require, being comparable to the orbital period. It may be possible to reduce this time by observing partial phase curves -- for example, starting before an eclipse and finishing after the night-side is observed. However, we suggest that the issues with instrumental systematics encountered for even a full phase curve with two anchoring eclipses in \citet{kempton2023reflective} imply that a partial phase curve will be impractical for constraining night-side emission.


\subsection{Observing Strategy}

We suggest three general strategies to spend 500 hours with JWST searching for atmospheres rocky exoplanets:

\begin{enumerate}
    \item Observing secondary eclipses with the F1500W filter only \citep{redfield2024report}; given 500 hours, five eclipses could be observed of each of the top 20 rocky planets sorted by signal-to-noise ratio in the F1500W bandpass. We suggest that these observations would be very susceptible to false positives mistaking high-albedo bare-rock surfaces for atmospheres, or to false negatives mistaking atmospheres with high emission for bare-rock surfaces.
    \item Observing secondary eclipses with the F1500W and F1280W filters \citep{ih2023constraining}; given 500 hours, five eclipses could be observed of each of the top 10 rocky planets sorted by signal-to-noise ratio in the F1500W bandpass. We suggest that these observations would be less degenerate than the first option, confidently identifying atmospheres with an intermediate amount of CO$_{2}$. However, many types of atmosphere would still be degenerate with bare-rock surfaces, and observing eclipses in two filters would be more time-consuming.
    \item Observing phase curves with the F1500W filter; given 500 hours, 1.2 orbits (including two eclipses) could be observed for each of 10 rocky planets selected for shorter orbital periods, spanning the range of equilibrium temperatures we model. An example sample would be LHS 1140 c (period 3.8 days), LTT 1445 A c (3.1 days), SPECULOOS-3 b (0.7 days), GJ 3929 b (2.6 days), GJ 1132 b (1.6 days), LHS 475 b (2.0 days), GJ 486 b (1.5 days),  LHS 3844 b (0.5 days), GJ 1252 b (0.5 days), and TOI-4527.01 (0.4 days). We suggest that this is the best method for unambiguous detections of atmospheres, as it does not rely on a particular atmospheric composition. Crucially, it avoids the degeneracy between atmospheres and high-albedo surfaces. Any atmosphere that transports enough heat to the night-side would be detectable in this way.
\end{enumerate}

We propose the third strategy as a way to provide unambiguous detections of atmospheres. As discussed above, this is a different conclusion about the optimal strategy to \citet{koll2019identifying}, which proposes eclipse-only photometry (the first strategy in our list) as a method \citep{mansfield2019identifying} to identify candidate atmospheres for subsequent follow-up with phase curve or spectroscopic observations. The primary reason for this difference is our use of the Bond albedo instead of the geometric albedo to model the temperature of bare-rock planets, resulting in a much stronger degeneracy than identified in \citet{mansfield2019identifying}.

It could be beneficial to follow up observations of phase curves in the F1500W filter with observations of phase curves or secondary eclipses in the F1280W filter, if there is weak global emission in the F1500W filter. If this were caused by a CO$_{2}$ absorption feature, the F1280W filter could detect stronger global emission as described in Section \ref{sec:results}. Another modification could be to additionally use observations from 5 to 12 $\mu$m with MIRI LRS, as these may be more optimal than the F1500W and F1280W filters for hotter planets emitting more at these shorter wavelengths. \citet{mansfield2024no} presented eclipse observations of GJ 486 b with this instrument, deriving a precise measurement of its brightness temperature but not detecting any clear spectral features. 

The three separate strategies described above could also be combined. For example, observing just two F1500W eclipses of each of the top 20 targets might identify targets with day-side emission significantly below the black body value, although with relatively large uncertainty. As we argue above that reducing the uncertainty on the day-side emission would very rarely produce a conclusive detection of an atmosphere, the planets with the lowest day-side emission could then be followed up with an observation of a phase curve. We stress again that it is perfectly plausible that the day-side atmosphere of a planet could emit at its black body temperature in the F1500W filter, while also emitting significantly from its night-side.  For example, the TRAPPIST-1 c simulation in Figure \ref{fig:compare_phase_emission_names} has day-side emission consistent with its black-body value; for weaker heat redistribution, more planets would also have day-side emission consistent with the black-body value (see the zero redistribution cases in Figure \ref{fig:compare_atmos_emission_names}). Other processes like thermal inversions or strong greenhouse warming could further increase the day-side emission for atmospheric compositions that we do not model. We reiterate, therefore, that targets for atmospheric detections should not be ruled out on the basis of black body day-side emission.

This strategy would be similar to the proposal of \citet{koll2019identifying} to use ``one to two eclipses with JWST \textup{[\,\dots]} confirmed by follow-up transit spectroscopy, eclipse spectroscopy,  or thermal phase curves''. We suggest that the difficulty of identifying spectral features from transit and eclipse spectroscopy of temperate rocky exoplanets \citep{lim2023trappist1b,lustig2023lhs475,wachiraphan2024ltt} promotes the use of phase curves for these follow-up measurements. \citet{ducrot2023combined} also suggests that a combination of broadband emission spectra with phase curve can provide robust detections of atmospheres on rocky planets, as they demonstrate that eclipse observations of TRAPPIST-1b in two photometric filters are not enough to rule out an atmospheric or a bare-rock scenario.



\section{Conclusions}\label{sec:conclusions}

We have presented simulations of JWST MIRI observations in the F1500W and F1280W filters, for a range of surfaces and atmospheres on a selection of the most readily observable rocky exoplanets. These have shown that the emission in the F1500W filter is highly degenerate between the surfaces and atmospheres. We suggest that it is more difficult than previously suggested to detect an atmosphere with observations in this filter alone. This is due to the increased range of possible emission values from bare-rock atmospheres that we model, and the similar scaling with temperature of bare-rock emission and emission from CO$_{2}$-dominated atmospheres. We suggest that no observation in this filter alone can unambiguously determine the presence or absence of an atmosphere, being prone to false positives or false negatives.

We also modelled an observational strategy using the difference in emission in the F1500W and F1280W filters, which is roughly uniform for the bare-rock surfaces, but can vary for different atmospheric compositions. A subset of atmospheres have detectably different emission in these bandpasses; these are generally atmospheres with enough CO$_{2}$ to create significant opacity in the F1500W bandpass, but not so much that there is also significant opacity in the F1280W bandpass. This technique would only be able to identify this subset of atmospheres with the correct spectral properties, with other types still indistinguishable from bare-rock surfaces with high albedo, and would also take longer than single-filter measurements.


We then simulated the detectability of night-side emission from phase curve observations in the F1500W filter. This showed that if $25\%$ of the heating on the day-side is redistributed to the night-side, almost every one of the simulated atmospheres would have detectable emission from its night-side given an observation of one full phase curve. We chose this fraction of $25\%$ as an example as this quantity is unknown for rocky planets in general; atmospheres with more or less redistribution would be more or less detectable. We suggest that finding night-side emission would be an unambiguous detection of an atmosphere. This technique would work for any atmosphere redistributing enough of its day-side heating regardless of composition. We suggest that a phase curve can provide a model-independent detection of an atmosphere, relying on the night-side emission that is simply a parameter of the fitted time-series model. Conversely, day-side emission will always be a model-dependent way to search for atmospheres, relying on complex expectations about which surface types could be feasible false positives, and on the emission simulated by more complex atmospheric models. 

Despite their theoretical ability to provide an unambiguous atmospheric detection, phase curves have limitations in reality. They are time-consuming for planets with long orbital periods, and are not necessarily easy to successfully execute due to issues with constraining instrumental systematics discussed in Section \ref{sec:discussion}. Night-side emission due to tidal dissipation could also provide a false positive detection of an atmosphere from a phase curve; we suggest above that this is unlikely to be detectable for these planets, but it must be considered when interpreting a phase curve.

Our conclusion that observations of eclipses are of very limited use for detecting atmospheres differs from \citet{mansfield2019identifying} and \citet{koll2019identifying}, which both proposed MIRI eclipses as a method to detect atmospheres on rocky planets (as candidates, in the case of \citet{koll2019identifying}). Section \ref{sec:discussion} identified that the primary cause of this difference was our use of the Bond albedo to calculate the temperature of bare-rock planets, rather than the use of the geometric albedo in \citet{mansfield2019identifying}. This produced an erroneously high eclipse depth for the bare-rock surfaces in \citet{mansfield2019identifying}, which was inherited by \citet{koll2019identifying} in its discussion of the bare-rock false positive. We also expanded on the range of surface types modelled in \citet{hu2012theoretical} and \citet{mansfield2019identifying}, which resulted in a stronger atmosphere-surface degeneracy (together with the corrected albedo). This motivated our argument that night-side emission is the only method by which an atmosphere could be unambiguously detected.

A strategy focused on phase curves would need to target planets with shorter orbital periods, which might restrict the distribution of the observed planets and the resulting statistical power of a test of the ``cosmic shoreline'' hypothesis. However, \citet{redfield2024report} state that the goal of the proposed 500 hour survey is to definitively identify which planets have atmospheres. We suggest that an observing strategy focused on observing phase curves of $\sim$10 planets would allow the unambiguous detections of atmospheres needed to meet this goal. This would still not be a trivial exercise given the effects of instrumental systematics on measurements of night-side emission. Future studies could investigate the sample sizes needed for statistical tests of hypotheses about the distribution of atmospheres on rocky planets. We suggest that any statistical tests, no matter how sophisticated, ultimately depend on unambiguous information content in each observation which could only be provided by phase curve observations.

\begin{acknowledgments}

M.H. is supported by Christ Church, University of Oxford. C.M.G. is supported by the Science and Technology Facilities Council [grant number ST/W000903/1]. T.L. acknowledges support by the Branco Weiss Foundation, the Alfred P. Sloan Foundation (AEThER project, G202114194), and NASA’s Nexus for Exoplanet System Science research coordination network (Alien Earths project, 80NSSC21K0593). H.N. was supported by the Clarendon Fund and the MT Scholarship Trust. C.F. acknowledges financial support from the European Research Council (ERC) under the European Union’s Horizon 2020 research and innovation program under grant agreement no. 805445. Support for this work was provided by NASA through the NASA Hubble Fellowship grant \#HST-HF2-51559.001-A awarded by the Space Telescope Science Institute, which is operated by the Association of Universities for Research in Astronomy, Inc., for NASA, under contract NAS5-26555. T.G.M. was supported by the SNSF Postdoc Mobility Grant P500PT\_211044

This research utilises spectra acquired by Raymond E. Arvidson, Melinda D. Dyar, Bethany L. Ehlmann, William H. Farrand, George Mathew, Jack Mustard, Carle M. Pieters, Hiroshi Takeda, and the Planetary Geosciences Lab (PSI) with the NASA RELAB facility at Brown University. 

The authors thank Edwin Kite, Eliza Kempton, and Jacob Bean for valuable feedback on an early draft of this manuscript. 
\end{acknowledgments}

\bibliography{sample631, additional_refs}{}
\bibliographystyle{aasjournal}

\end{document}